\documentclass[graybox]{svmult}

\usepackage{type1cm}        
\usepackage{makeidx}         
\usepackage{graphicx}        
\usepackage{multicol}        
\usepackage[bottom]{footmisc}

\usepackage{braket}

\usepackage{newtxtext}       %
\usepackage[varvw]{newtxmath}       

\makeindex             

\begin{document}

\title*{Quantum Correlations in Neutrino and Neutral Meson Oscillations}
\author{Subhashish Banerjee\orcidID{0000-0002-7739-4680}}
\institute{Subhashish Banerjee \at Indian Institute of Technology Jodhpur, Rajasthan \email{subhashish@iitj.ac.in}}
%
%
\maketitle


\abstract{We discuss the impact of ideas of open quantum systems and quantum information to various facets of neutrino and neutral meson oscillations. These oscillations are characterized by a number of quantum correlations, both spatial as well as temporal. For neutrinos, the correlations are shown to be simple functions of the product of neutrino survival and oscillation probabilities. The quantum correlations in the neutral mesons are seen to be non-trivially different from their stable counterparts.}

\section{Introduction}
The theory of open quantum systems addresses the issues of damping and dephasing by positing that all practical systems interact with their surrounding environments, thereby classifying them as 'open' systems \cite{SBbook}. Quantum optics was among the first fields to apply and test the concept of open quantum systems \cite{wl73}. The application of this theory to other fields was significantly advanced by the work of \cite{cl83}, \cite{wz93}, and others. The recent increase in interest in open quantum systems is driven by remarkable progress in manipulating quantum states of matter, and in encoding, transmitting, and processing quantum information \cite{turch, myatt, haroche}. Understanding and controlling environmental impacts is crucial for all these advances. This enhances the importance of open system notions for quantum computation and information.

In recent years, there has been an active intersection between ideas of open quantum systems and quantum information with sub-atomic physics. Here, we will focus on this issue. In particular, we will study neutrino and neutral meson oscillations in this framework. Pauli proposed the existence of neutrinos to bring about energy and momentum conservation in beta decay~\cite{kim}. Neutrinos are now understood to come in three flavours~$\nu_{e}, \nu_{\mu}, \nu_{\tau}$, that is, electron, muon, and tau neutrinos. They oscillate because they have non-zero mass~\cite{kim}. Oscillating neutrinos fundamentally involve three flavours but can often be approximated as two-flavour scenarios.
Prominent issues related to neutrinos are the neutrino mass hierarchy, CP (charge-parity) violation, among others. The investigation of these issues are the primary focus of the current neutrino experiments including the (long-baseline) accelerator ~\cite{Dune}.

Neutral mesons $K^0, B^0_d, B^0_s$ are unstable massive systems undergoing oscillations. The study of single and decaying correlated mesons provide an interesting platform for open system ideas to decaying quantum systems. Further, correlated neutral mesons facilitate the interplay of quantum correlations and CP-symmetry.


 
 The article is laid out in the following manner. Section II introduces open quantum systems and their connection to quantum channels. Section III provides a brief introduction to several measures of quantum correlations, both spatial and temporal, that will be subsequently used on neutrino and meson oscillations. Quantum correlation in neutrino oscillations are then discussed, followed by neutral meson systems. We then make our conclusions. 



\section{Open Quantum Systems}
\label{sect2}
Here, we will make explicit what we mean by open quantum systems~\cite{SBbook}. To illustrate the formalism, we will provide simple examples of open systems, of  relevance to quantum information. Following this, we will connect open system ideas to quantum information. 
Consider the Hamiltonian of the total (closed system):
\begin{equation}
    H = H_S + H_R + H_{SR},
    \label{4a}
\end{equation}
where $S$ represents the system, $R$ denotes the reservoir (bath), and $SR$ signifies the interaction between them.


The given Hamiltonian of the system-reservoir follows unitary evolution as:
$
\rho (t) = e^{-{i \over \hbar}Ht} \rho (0) e^{{i \over 
\hbar}Ht}. 
$
We focus on the reduced dynamics of the system $S$, while accounting for its environmental influence. This is accomplished by tracing over the reservoir degrees of freedom to obtain the reduced dynamics, which is {\em non-unitary}:
$
\rho^S (t) = \rm{Tr}_R[\rho (t)].
$
Open quantum systems are commonly categorized into two main types: (a) Quantum non-demolition  (QND), where $[H_S, H_{SR}] =
0$, leading to decoherence without dissipation \cite{brag75, cave80}
  and (b) Quantum  dissipative  systems, where  $[H_S,  H_{SR}]  \ne  0$ resulting in decoherence accompanied by dissipation \cite{cl83,gsi88}.


In quantum information theory, the noise generated by a QND open system is referred to as a ``phase damping channel'', whereas the noise from a dissipative (Lindblad) evolution is termed a ``(generalized) amplitude damping channel''.


\subsection{System-Reservoir Initial Conditions:}


The evolution of the open system begins with the system and the reservoir either initially uncorrelated or correlated.

(A). Separable  Initial Condition: It is assumed that the system and the environment (reservoir) are initially independent of each other \cite{fv63, cl83}. Under these conditions, the initial density matrix can be expressed as $
\rho (0) = \rho^S (0) \otimes \rho^R(0) ,
$ 
where $\rho^S  (0)$ and 
$\rho^R (0)$ represents the system's and reservoir's initial density matrices, respectively.

(B). Non-Separable   Initial Condition: in many scenarios, the system and the reservoir are interconnected as parts of a larger system, and their interaction is beyond our control. This situation prompts the use of a specific class of initial conditions known as `generalized initial conditions' \cite{ha85, gsi88}. A broad category of these initial conditions can be expressed as $
\rho_0 = \sum\limits_j O_j \rho_{\beta} O'_j ,
$
where
$
\rho_{\beta} = Z^{-1}_{\beta} \exp (-\beta H)
$ is the canonical density matrix that describes the equilibrium state of the interacting system in the presence of a time-independent potential $ V
$, with $ Z^{-1}_{\beta} $ being the partition function.  Here $ \beta = (
k_B T)^{-1}  $, where $  T $ is the equilibrium temperature of the interacting system. The operators $O_j,  O'_j$ act solely on the system's coordinates, leaving the reservoir's coordinates (environment) unchanged, though they can be selected arbitrarily otherwise.


{\it Time scales associated with the Open System Evolution:} Various time scales characterize the open system evolution, with the most notable being: \\ (a). Scale related to the system's natural frequency \\ (b). Relaxation time scale governed by the strength of the $S-R$ coupling \\ (c). Reservoir correlation time (memory time), linked to the high-frequency cutoff in the reservoir's spectral density and the time scale related to the reservoir temperature, which reflects the relative significance of quantum effects compared to thermal effects


{\it Dissipative form of evolution equation: Lindblad form:} A widely used form of the evolution equation for the reduced density matrix of a system, applicable when the system and interaction Hamiltonians do not commute, is the Lindblad master equation. This equation is based on the physical assumptions of the {\em Born} (weak coupling), {\em Markov}  (memoryless), and {\em Rotating Wave Approximation} (rapid system dynamics relative to the relaxation time).

A common prototype for the Lindblad evolution would be the decay of a two-level system $H_S = \frac{1}{2}  \hbar \omega_0 \sigma_z$, interacting
with a radiation field (bath) in the weak Born-Markov, rotating wave
approximation. 
In the interaction picture (ignoring the Lamb shift terms), the Lindblad master equation for $\rho^S(t)$ is
\begin{equation}
\hspace*{-1cm}{d   \over   dt}\rho^S(t) =  \gamma_0\left[ (N_{th}+1)\chi_1(t) +
  N_{th} \chi_2(t)\right], \label{4b} 
\end{equation}
where
\begin{align}
    \chi_1(t)&=\left[\sigma_-  \rho^S(t)  \sigma_+  -  {(1/2)}\sigma_+  \sigma_-
\rho^S(t) - {(1/2)}\rho^S(t) \sigma_+ \sigma_- \right],\nonumber\\
\chi_2(t)&=\left[  \sigma_+  \rho^S(t)
\sigma_-  - {(1/ 2)}\sigma_-  \sigma_+ \rho^S(t)  -  {(1/2)}
\rho^S(t)  \sigma_- \sigma_+  \right].\nonumber
\end{align}

Here, $\gamma_0$ is spontaneous emission rate 
$
\gamma_0 = {4 \omega^3 |\vec{d}|^2 \over 3 \hbar c^3},
$
and   $\sigma_+$,  $\sigma_-$:  standard   raising  and
lowering operators, respectively given by
$
\sigma_+ = |1 \rangle \langle 0| =  {1 \over 2}
\left(\sigma_x + i \sigma_y \right);~~~
\sigma_- = |0 \rangle \langle 1| = {1 \over 2}
\left(\sigma_x - i \sigma_y \right).  
$In the RHS of Eq.\,(\ref{4b}), the first parenthesis, involving
$ \gamma_0   (N_{th}   +   1)$, describes both spontaneous emission  ($\gamma_0$) and thermal emission ($\gamma_0 N_{th}$). The second parenthesis, which includes
$\gamma_0 N_{th}$, represents thermal absorption.
Additionally, $N_{th} = {1 \over e^{{\hbar \omega \over k_B T}} - 1} 
$ represents the Planck distribution, which gives the count for thermal photons at the frequency  $\omega$. 


The two-level density matrix using Pauli operators can be expressed as
\begin{eqnarray}
\rho^S(t) &=& \left(\frac{I + \langle \vec{\sigma}(t) \rangle . 
\vec{\sigma}}{2}\right) = \left(
\begin{array}{ll} \alpha_+(t) &
\langle \sigma_- (t) \rangle   \\     ~     &    ~     \\
\langle \sigma_+ (t) \rangle 
 ~~~~~ & \alpha_-(t) 
\end{array}\right),
\label{4c}
\end{eqnarray} 
where $\alpha_{\pm}(t)=(1/2)\left(1 \pm \langle \sigma_z (t)
\rangle\right)$.
Using Eq.\,(\ref{4c}), we can readily derive Eq.\,(\ref{4b}).
This allows us to solve for the Bloch vectors, resulting in
\begin{eqnarray}
\langle \sigma_x (t) \rangle &=& e^{-{\gamma_0 \over 2}(2N_{th} + 1)t}
\langle \sigma_x (0) \rangle, \label{4h} 
~~\langle \sigma_y (t) \rangle = e^{-{\gamma_0 \over 2}(2N_{th} + 1)t}
\langle \sigma_y (0) \rangle, \label{4i} \nonumber\\
\langle \sigma_z (t) \rangle &=& e^{-\gamma_0 (2N_{th} + 1)t} \langle 
\sigma_z (0) \rangle - {1 \over (2N_{th} + 1)} \left(1 - e^{-\gamma_0 
(2N_{th} + 1)t} 
\right). \label{4j} 
\end{eqnarray}

\subsection{Connection to quantum noise processes}


Now, we seek to relate the above findings to common noisy channels: In what ways can quantum computing be influenced by environmental factors?
In operator-sum representation, the action of a superoperator $\mathcal{O}$ due
to environmental interaction is
$$
\label{eq:kraus}
\rho \longrightarrow {\mathcal{O}}(\rho) = 
\sum_k \langle e_k|\mathcal{U}(\rho \otimes |f_0\rangle\langle f_0|)\mathcal{U}^{\dag}|e_k\rangle
= \sum_j E_j \rho E_j^{\dag}.
$$ 
This expression describes the unitary operator $\mathcal{U}$, which accounts for the free evolution of the system with initial state $\rho$, the environment, and their interaction. Here, $|f_0\rangle$ denotes the initial state of the environment, and $\{|e_k\rangle\}$ represents a basis for the environment. It is assumed that the initial states of the system and environment are separable.
 $E_j  \equiv \langle  e_k|\mathcal{U}|f_0\rangle$ implements the dynamical mapping; 
and a partition of  unity: $\sum_j  E_j^{\dag}E_j  = \mathcal{I}$.
Any  transformation representable  as operator-sum  is  a completely
positive (CP) map  \cite{nc00}. This method is referred to as the quantum operations formalism \cite{sting55,sud61,kr83}.


\section{Measures of Quantum Correlations}
\label{Sect3}
Quantum correlations are a {\it many-faceted} entity. Some of the prominent ones are discussed below.

\subsection{Bell's inequality}
This measures the non-local characteristics of the system. If $\rho$ is the density matrix for a pair of qubits, the correlation matrix $T$ with its matrix elements can be expressed as $T_{mn}=Tr\left[\rho(\sigma_m\otimes \sigma_n)\right]$. The eigenvalues
$u_i$ ($\text{where, }i=1,2,3$) of the matrix $T^{\dagger}T$ quantify the Bell-CHSH inequality as $M(\rho)<1$~\cite{Horodecki_1996}. Here, $M(\rho)=\max(u_i+u_j)\ (i\neq j)$.

	
\subsection{Entanglement: Concurrence}
This is the most well known measure of quantum correlation. 
The concurrence ($C$) is an entanglement measure for a mixed state $\rho$ of two qubits and is defined as $ C=\max(\lambda_1-\lambda_2-\lambda_3-\lambda_4,0)$. 
To compute concurrence, first, find the eigenvalues of the matrix: 
$\sqrt{\rho}(\sigma_y\otimes \sigma_y)\rho^*(\sigma_y\otimes \sigma_y)\sqrt{\rho}$ and take their square roots. These square roots arranged in decreasing order are 
$\lambda_i$. Here, $\rho^*$ is calculated on the two-qubit computational basis ~\cite{Wootters_1998}. 
The entanglement of formation is equivalent to the concurrence in a two-qubit system and can be expressed as a monotonic function of the concurrence $C$ given by
$E_F=-\frac{1+\sqrt{1-C^2}}{2}\log_2(\frac{1+\sqrt{1-C^2}}{2})-\frac{1-\sqrt{1-C^2}}{2}\log_2(\frac{1-\sqrt{1-C^2}}{2}).$

\subsection{Teleportation Fidelity:}
Teleportation gives practical significance to the concept of entanglement.
When $F_{\max}>\frac{2}{3}$, teleportation becomes possible. 
$F_{\max}$ can be determined using the eigenvalues
$\{u_i\}$ of $T^\dagger T$.
$F_{\max}=\frac{1}{2}\left(1+\frac{1}{3}N(\rho)\right)$ where $N(\rho)=\sqrt{u_1}+\sqrt{u_2}+\sqrt{u_3}$~\cite{Horodecki_1996}. An inequality involving $M(\rho)$ and $F_{\max}$ is
$
F_{max}\ge \frac12\left(1+\frac13 M(\rho)\right)\ge\frac23\ {\rm if}\ M(\rho)>1.
$

\subsection{Measurement  induced  disturbance (QMID)}
QMID measures the extent of quantum correlation between the bipartite states shared by two parties, say Alice and Bob.
For the given  $\rho'_{A,R}$, if $\rho'_A$ and $\rho'_R$ are the reduced density matrices, then  the mutual information that quantifies the correlation between    Alice and Rob is    
$	
I=S(\rho'_A)+S(\rho'_R)-S(\rho'_{A,R}),  
$ 
where, $S(.)$  is the von Neumann  entropy. If $\rho'_A=\sum_i  \lambda_A^i  \Pi_A^i$  and $\rho'_R=\sum_j \lambda_R^j \Pi_R^j$ denotes the spectral decomposition of $\rho'_A$  and $\rho'_R$,  respectively, then the  state $\rho'_{A,R}$ after   measuring   in    joint   basis   $\{\Pi_A,\Pi_R\}$ is   
$
\Pi(\rho^\prime_{A,R})=\sum_{i,j}(\Pi_A^i\otimes\Pi_R^j)\rho^\prime_{A,R} (\Pi_A^i\otimes\Pi_R^j).  
$
QMID~\cite{Luo_2008} is         
$
M(\rho^\prime_{A,R})=I(\rho^\prime_{A,R})-I(\Pi(\rho^\prime_{A,R}))
$
is  a measure  of quantumness of  the correlation.


\subsection{Discord} 
Quantum discord (QD)~\cite{Zurek_2001, Vedral_2001} quantifies the disparity between total and classical correlations. It is derived from two distinct methods of extending classical mutual information into the quantum domain: one adheres to conventional mutual information, and the other represents a quantum interpretation of conditional entropy influenced by the measurement process.
It is defined as 
$
    \mathcal{J}(\rho_P|\rho_C) = \mathcal{S}(\rho_P) - \mathcal{S}(\rho_P|\Pi_i^C) 
$
where $\mathcal{S}(\rho_P|\Pi_i^C)$ is the quantum conditional entropy, defined with respect to a set of projective measurements $\left \{\Pi_i \right\}$ performed on the state $\rho_C$. 
QD is  
$
    \mathcal{D}(\rho_P|\rho_C)= \mathcal{I}(\rho_P:\rho_C) - \max_{\Pi^C_i} \mathcal{J}(\rho_P|\rho_C). 
$ A conceptually similar though computationally easier measure is the geometric discord $D_G$~\cite{Brukner_2010}. An optimal version of QMID can be viewed as QD. QMID characterizes quantum correlations using spectral projections of measurement operators without any optimization and makes it relatively straightforward to compute. In contrast, QD requires optimization over all possible projectors to calculate the correlations between states, resulting in high computational complexity.



\subsection{Leggett-Garg Inequality:}
This epitemizes the temporal correlations~\cite{LG_1985}. It is based on the assumptions: (a) Macroscopic realism (MR): A macroscopic system with two or more distinctly different states will always be in one of these states at any given time; (b) Non-Invasive measurability (NIM): In principle, it allows to perform the measurement without disturbing the future dynamics of the system. A central quantity here is the two-time correlation and involves dichotomic observable $Q = \pm 1$.
\begin{center}
	\includegraphics[scale = 0.35]{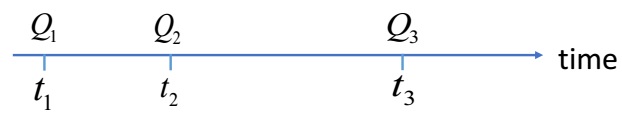}
\end{center}
The two-time corrlation functions have the form  $C_{ij} = \frac{1}{N} \sum\limits_{q=1}^{N}  Q^{q}_{i} Q^{q}_{j}$, where $-1 \le C_{ij} \le 1$. $C_{ij} = \pm 1$ implies perfectly (anti-)correlated case, while $C_{ij} = 0$ implies no correlation. The simplest form of the Leggett-Garg Inequality is:
$
	K_3 = C_{12} + C_{23}  - C_{13} =  \langle Q_1 Q_2 \rangle + \langle Q_2 Q_3 \rangle -  \langle Q_1 Q_3 \rangle. 
$
It can be easily shown that $-3 \le K_3 \le 1$.

\section{Quantum correlations in neutrino oscillations}

Quantum correlations serve a vital role
in understanding \& harnessing the power of quantum mechanics across various natural systems. Traditionally, studies of quantum mechanics foundations have centered on optical and electronic systems, where different measures of quantum correlations are well-established. However, recent technical advances in high-energy physics, particularly in neutrino oscillation experiments, open new avenues for exploring these correlations in neutrinos. Here, we are interested in examining various aspects of quantum correlations in neutrino oscillation, including non-locality, entanglement, and weaker measures like discord.

The neutrino system is especially intriguing because it experiences minimal decoherence compared to other particles commonly used in quantum information processing. This characteristic makes neutrinos a valuable platform for exploring the foundations of quantum mechanics.

The coherent time evolution of neutrino flavour eigenstates indicates that a flavour state $\nu_\alpha$ ($\alpha=e,\mu,\tau$) can be represented as a linear combination of its mass eigenstates $\nu_j$ ($j=1,2,3$). Neutrino oscillations can be effectively described using transition probabilities between flavour states. The three flavor states ($\nu_\alpha$) of neutrinos are eigenstates of weak interaction that are detectable in the laboratory and can be mixed via a $3\times3$ unitary matrix to form the three mass eigenstates $\nu_j$.
These mass eigenstates propagate in space over time, resulting in the time-evolved flavor state. Neutrino oscillations occur only if the three corresponding masses, $m_1, m_2$, and $m_3$, are non-degenerate. Among the three mass-squared differences $\Delta_{ij}
= m_i^2 - m_j^2$ (where $i,j = 1,2,3$ with $i > j$), only two are independent. Oscillation measurements show that $\Delta_{21} \approx 0.03 \times \Delta_{32}$,
suggesting $\Delta_{31} \approx \Delta_{32}$.
\vspace{0.3cm}


In general, neutrino oscillations are described by the full three-flavour oscillation equations. However, in certain scenarios, these equations can be simplified to an effective two-flavour model by setting one or both of the small parameters, $\Delta_{21}/\Delta_{32}$ and $\theta_{13}$, to zero. For instance, in long-baseline accelerator experiments, these parameters can often be ignored for leading-order calculations. This simplification reduces the problem to two-flavour mixing, where flavour neutrinos  $\nu_\alpha$
and $\nu_\beta$ mix to produce two mass eigenstates, $\nu_j$ and $\nu_k$. In this simplified model, oscillations are characterized by a single mixing angle
$\theta$ ($\equiv \theta_{23}$) and a single mass-squared difference $\Delta$ ($\equiv \Delta_{32}$). The flavour and mass eigenstates are related through
$2\times2$ rotation matrix, $U(\theta)=\left(\begin{array}{cc} \cos (\theta) & \sin( \theta) \\ - \sin (\theta) & \cos (\theta) \end{array}\right)$, as

\begin{align}
\left(\begin{array}{c} \nu_{\alpha} \\ \nu_{\beta} \end{array}\right) = 
U(\theta) 
\left(\begin{array}{c} \nu_j \\ \nu_k \end{array}\right)\,,
\end{align}
Thus, each flavour state $\ket{\nu_\alpha}$ represents a superposition of mass eigenstates: $
\ket{\nu_{\alpha}} = \sum_j U_{\alpha j} \ket{\nu_j}\
$, where $\alpha = \mu$ or $\tau$ and $j = 2,3$.
The time evolution of the mass eigenstates $\ket{\nu_{j}}$ is
$
\ket{\nu_{j}(t)} = e^{-i E_{j} t} \ket{\nu_{j}}\,,
$
where $\ket{\nu_{j}}$ are the mass states at time $t=0$. 
Therefore,
\begin{align}
\ket{\nu_{\alpha} (t)} = \sum_j e^{-i E_j t} U_{\alpha j} \ket{\nu_j}\,.
\label{numass}
\end{align}
It is also possible to project the evolving flavor neutrino state $\ket{\nu_{\alpha}}$ onto the flavor basis as 
\begin{equation}
\ket{\nu_{\alpha}(t)} = \tilde{U}_{\alpha \alpha} (t)\ket{\nu_{\alpha}} + \tilde{U}_{\alpha \beta} (t)\ket{\nu_{\beta}}\,,\hspace{0.2cm}\text{where
}\hspace{0.1cm} |\tilde{U}_{\alpha \alpha}(t)|^2+|\tilde{U}_{\alpha \beta}(t)|^2 = 1\,
\label{t2}
\end{equation}
and $\ket{\nu_{\alpha}}$ is the flavour state at time $t=0$. 
The flavour eigenstates are clearly explained in the framework of quantum mechanics, where all detectable neutrinos are ultra-relativistic. This allows us to establish a correspondence with two-qubit states by using the occupation number of neutrinos as~\cite{Blasone_2008, Blasone_2009}
\begin{equation}
\ket{\nu_{\alpha}}\equiv \ket{1}_{\alpha} \otimes \ket{0}_{\beta}\equiv \ket{10},\hspace{.16cm}
\ket{\nu_{\beta}}\equiv \ket{0}_{\alpha} \otimes \ket{1}_{\beta}\equiv \ket{01}\,.
\label{t3}
\end{equation}
Using Eq.\,(\ref{t3}) in Eq.\,({\ref{t2}}), we get 
\begin{equation}
\ket{\nu_{\alpha}(t)} = \tilde{U}_{\alpha \alpha} (t)\ket{1}_{\alpha} \otimes \ket{0}_{\beta} + \tilde{U}_{\alpha \beta} (t) \ket{0}_{\alpha} \otimes \ket{1}_{\beta}\,
\label{flavmode}
\end{equation}
where,
\begin{eqnarray}
	\tilde{U}_{\alpha \alpha} (t) = \cos^2( \theta) e^{-i E_2 t} + \sin^2( \theta) e^{-i E_3 t}; 
	\tilde{U}_{\alpha \beta} (t) = \sin (\theta) \cos( \theta) (e^{-i E_3 t} - e^{-i E_2 t}) \nonumber\\ 
\end{eqnarray}
implies that $\ket{\nu_{\alpha}(t)} $ exhibits an entangled state. The survival and oscillation probabilities of the initial state $\ket{\nu_\alpha}$ are given by $P_{sur}=|\tilde{U}_{\alpha \alpha} (t) |^2=\frac{1}{4}(3 + \cos(4\theta) + 2\cos( \phi) \sin^2 (2\theta))$ and $P_{osc}=|\tilde{U}_{\alpha \beta} (t)|^2=\sin^2 (2\theta) \sin^2\left(\phi/2\right)$, respectively, where, $\phi=\frac{\Delta t}{2 E}$. Here, $P_{sur}+P_{osc}=1$.
Since the density matrix $\rho_{\alpha}(t)=\ket{\nu_{\alpha}(t)}\bra{\nu_{\alpha}(t)}$ incorporates the mixing angle ($\theta$) and the mass-squared difference ($\Delta$), which are known, various quantum correlation measures can now be evaluated using this matrix~\cite{SB_2016, SB_2015}.
\begin{enumerate}
    \item Non-locality:
\begin{align}
M(\rho)& = 1 + \Big[3 + \cos(4\theta) + 2\cos( \phi) \sin^2 (2\theta)  \Big] \sin^2 (2\theta) \sin^2\left(\phi/2\right)\nonumber\\
&= 1 + 4 P_{sur}P_{osc}.
\end{align}
 $M(\rho)$ is tied up with neutrino oscillation probabilities. When $\theta=0$ (no mixing), $M(\rho)=1$.
\item Concurrence:
\begin{align}
C &= \sqrt{(3 + \cos(4\theta) + 2\cos (\phi) \sin^2 (2\theta) )}
		\sin (2\theta) \sin \left(\phi/2\right)\nonumber\\
  &= 2\sqrt{P_{sur} P_{osc}}.
\end{align}
In case of no oscillation, there is no entanglement. Also, $M(\rho)=1+C^2$.
\item Geometric discord:
\begin{align}
D_{G}(\rho)&= \frac{2}{3} \Big[3 + \cos( 4 \theta) + 2 \cos( \phi) \sin^2( 2 \theta) \Big]\sin^2 (2\theta) \sin^2 \left(\phi/2\right)\nonumber\\
&= \frac{8}{3}P_{sur}P_{osc}.
\end{align}
Further, $ D_{G}(\rho) \neq 0$ for $P_{osc} \neq 0$. 
\item Teleportation fidelity:
\begin{align}
F_{\max} &= \frac{1}{3}\left[2 +
		(3+\cos(4\theta) + 2 \cos (\phi)\sin^2 (2\theta))^{1/2} \sin(2 \theta) \sin \left(\phi/2\right) \right]\nonumber\\
  &= (2/3)(1+ \sqrt{P_{sur}P_{osc}}). 
\end{align}
For nonzero $P_{osc}$, $F_{\max} > 2/3$. 
\end{enumerate}

Thus, we can see that Bell's inequality is consistently violated, demonstrating the non-local nature of neutrino evolution. The teleportation fidelity remains above 
2/3, which obeys the typical relationship between Bell's inequality violation and teleportation fidelity observed in electronic and photonic systems. These quantum correlations, which are functions of the product of neutrino survival and oscillation probabilities, can exhibit classically forbidden values for non-zero oscillation probabilities. The strength of these correlations is closely linked to the neutrino mixing angle. This analysis can be extended to the case of three flavor neutrino oscillations and has been used to study both spatial and temporal correlations~\cite{SB_2019, SB_2020}, neutrino mass degeneracy problem~\cite{SB_2018}, CP-violation in neutrino oscillations~\cite{Bouri:2024kcl}, neutrino oscillations in curved space-time~\cite{SB_BM_2018}, among others.

The various quantum correlations used in this work are indeed helpful for studying bipartite entanglement in oscillating neutrinos. A comprehensive analysis is made in Ref.\,\cite{KumarJha:2020pke}, where quantum correlations such as three-tangle and three-$\pi$ have been used to illustrate genuine tripartite entanglement in oscillating neutrinos. Additionally, since qutrits have higher dimensions than qubits, quantum correlations such as generalized concurrence (an extension of concurrence in a higher dimension) have been quantified by treating neutrinos as qutrits \cite{Jha:2022yik}. Inspired by these findings, implementing neutrino entanglement on the IBM quantum computer appears promising for future studies of quantum correlations associated with neutrino oscillations in realistic situations which may be difficult to treat otherwise \cite{Jha:2021itm}. 

\section{Quantum correlations in neutral meson systems}

In this section, we analyze various widely recognized quantum correlation measures,  including Bell's inequality violations, teleportation fidelity, concurrence, and geometric discord, for the correlated $B\bar B$ and $K\bar K$ systems. Additionally, we explore the interplay between these measures~\cite{AK_SB_2013, AK_SB_MK_2016}. $B$ factories, electron-positron colliders designed specifically for studying the production and decay of $B$ mesons, and $\phi$ factories, which serve the same purpose for $K$ mesons, offer an ideal testing ground. The collider energy is set to the $\Upsilon$ resonance in $B$ factories, which initiates the process $e^+e^-\to\Upsilon$. The $\Upsilon$ then decays into $b\bar b$, which subsequently form $B_q \bar{B}_q$ ($q=s,d$) pair through hadronization. This process occurs almost instantaneously, while the $B$ mesons separate and decay for a much longer period. An important aspect of these systems for studying correlations is the oscillation of bottom and strangeness flavours, leading to $B\bar B$ and $K\bar K$ oscillations. Since these systems are unstable and decaying, it is essential to investigate quantum correlations in the context of these unstable, decaying $B\bar B$ and $K\bar K$ systems. 

Using the state of the two-particle decaying system at time $t$~\cite{AK_SB_MK_2016}, we have the following quantum correlations in neutral mesons
\begin{enumerate}
    \item \textit{Non-locality}:
	$M(\rho)=(1+e^{-4 \lambda t}).$ Here $\lambda$ characterizes the decoherence, that in turn arises due to the irreversibility caused by the decay.
 \item \textit{Concurrence}:
	$
	    C=e^{-2\lambda t}
	$. The corresponding entanglement of formation is 
	$
		E_F=-\frac{1+\sqrt{1-C^2}}{2}\log_2(\frac{1+\sqrt{1-C^2}}{2})-\frac{1-\sqrt{1-C^2}}{2}\log_2(\frac{1-\sqrt{1-C^2}}{2}).
	$ Also, $M(\rho)=1+C^2$.
\item \textit{Geometric discord}: $ D_{G}(\rho)=M(\rho)/3.$
\item \textit{Teleportation fidelity}: $
F_{\max} = \frac{1}{12}\Big[6+2e^{-2\lambda t} +\sqrt{2}\sqrt{\alpha-\sqrt{\beta}} + \sqrt{2}\sqrt{\alpha+\sqrt{\beta}}\Big],
$ where
	$\alpha = 1 + \cosh(4 \lambda t)-\sinh(4 \lambda t),$ and 
	$\beta = 3 - 2\alpha + \cosh(8 \lambda t)-\sinh(8 \lambda t).$
\end{enumerate}

To incorporate the effects of decay in the systems under investigation, the various correlations are modified by the probability of survival of the particle pair up to a given time, which is expressed as $e^{-2\Gamma t}$, where $\Gamma$ represents the meson decay width. For the $K$ meson, the decay width is given by $\Gamma=\frac{1}{2}(\Gamma_S + \Gamma_L)$, with $\Gamma_S$ and $\Gamma_L$ being the decay widths of the short and long neutral kaon states, respectively; its value is $5.59 \times 10^{9}\,\rm s^{-1}$ ~\cite{PDG_2014}. 
The decay widths for $B_d$ and $B_s$ mesons are $6.58 \times 10^{11}\, \rm s^{-1}$ and $6.61 
	\times 10^{11}\, \rm s^{-1}$, respectively~\cite{HFAG_2012}. 
For the $K$ meson system, the decoherence parameter 
 $\lambda$ determine by the KLOE collaboration through the interference between initially entangled kaons and their decay products in the channel $\phi \to K_S K_L \to \pi^+ \pi^- \pi^+ \pi^-$~\cite{KLOE_2006}. The value of $\lambda$ is constrained to be no more than $1.58 \times 10^9\, {\rm s^{-1}}$ at the $\rm 3 \sigma$ level. For $B_d$ meson systems, the decoherence parameter is typically evaluated through time-integrated dilepton events. The value of $\lambda$ for $B_d$ mesons is calculated by the measurement of $R_d$, the ratio of the total same-sign to opposite-sign dilepton rates in the decays of coherent $B_d - \bar{B}_d$ coming from the $\Upsilon (4S)$ decays; the upper bound is $2.82 \times 10^{11}\, {\rm s^{-1}}$ at $\rm 3 \sigma$~\cite{Grimus_2001}. 
 In~\cite{AK_SB_SU_2015}, several methods were proposed to calculate $\lambda$ in the $B$ meson systems. For $B_s$ mesons, due to the lack of experimental evidence for $\lambda$,  we assume it to be zero in this study.

\begin{figure} 
	\centering
	\begin{tabular}{ccc}
		\includegraphics[width=30mm]{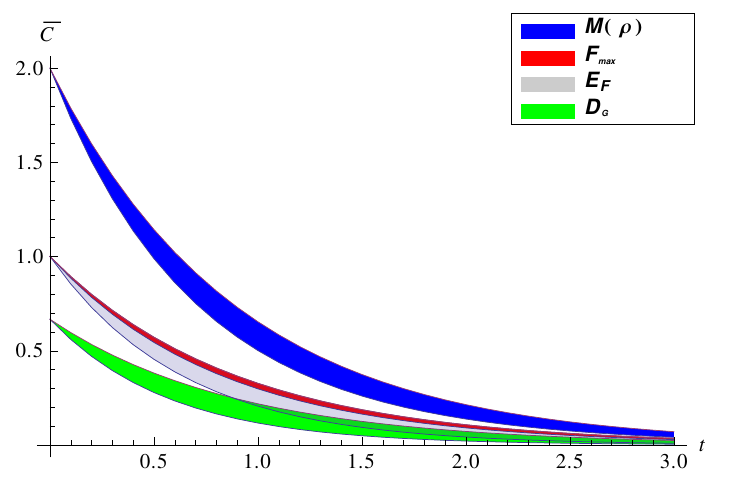}&
		\includegraphics[width=30mm]{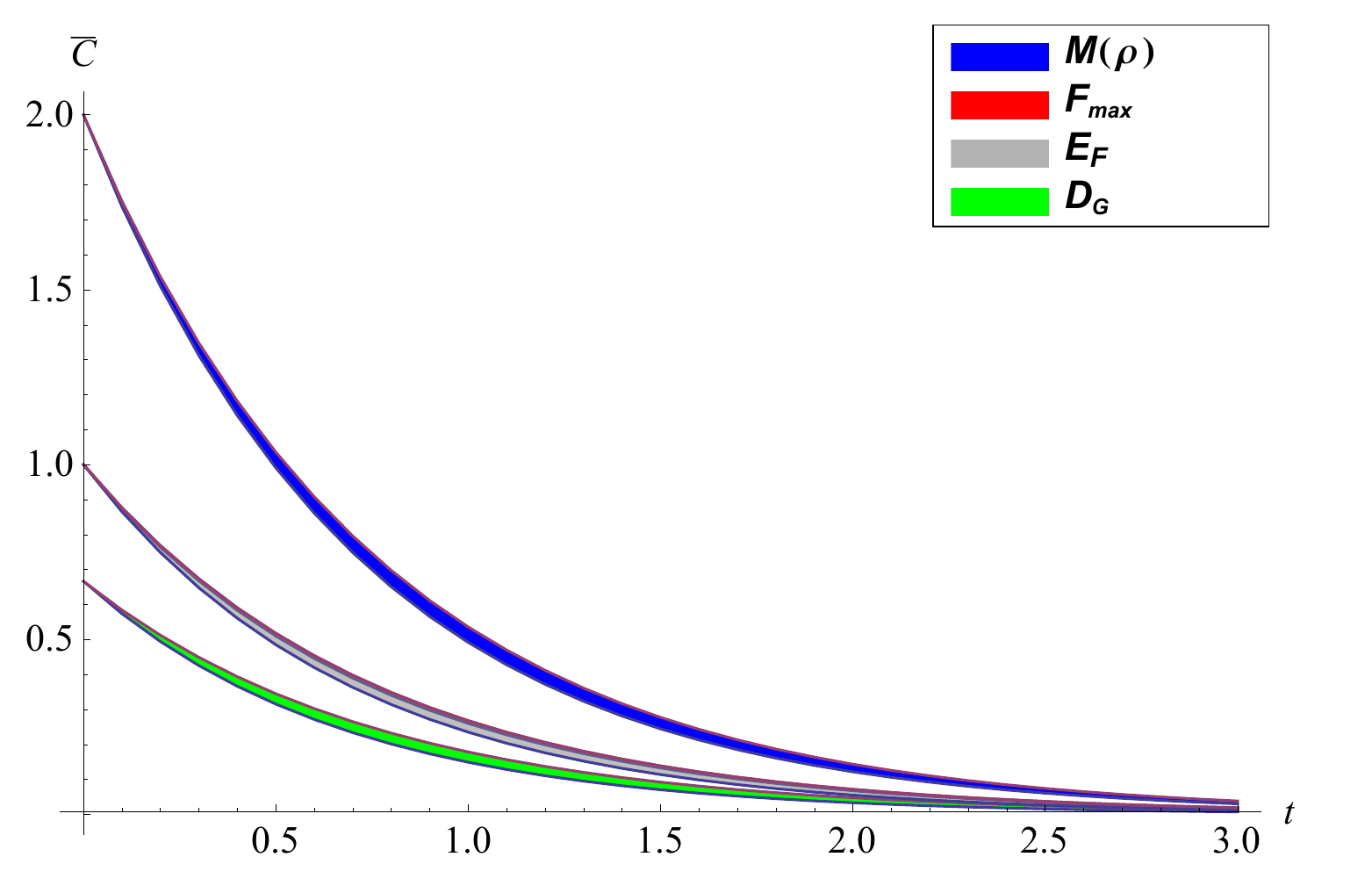}&
		\includegraphics[width=30mm]{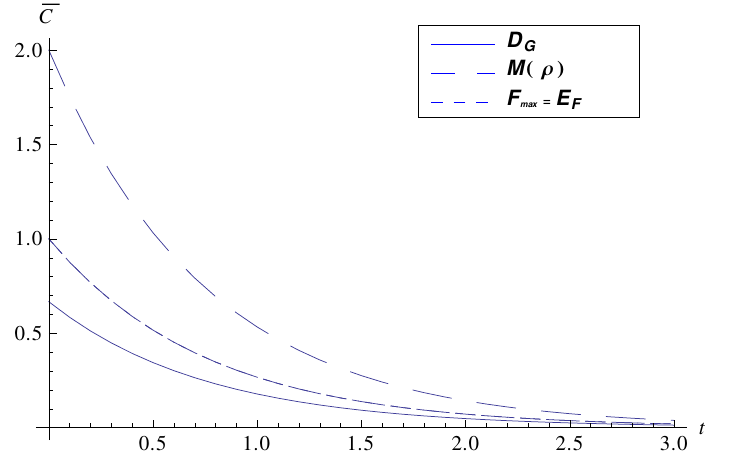}
	\end{tabular}
	\caption{Average correlation measures i.e., the various measures modulated by the exponential factor $e^{-2\Gamma t}$, are plotted as a function of time $t$. The left, middle, and right plots show the correlations for $K\bar{K}$, $B_d\bar{B_d}$ and $B_s\bar{B_s}$ pair created at $t=0$, respectively. The four correlation measures displayed from top to bottom are $M(\rho)$ (Bell's inequality; blue band), $F_{\max}$ (teleportation fidelity; red band), $E_F$ (entanglement of formation; grey band) and $D_G$ (geometric discord; green band). For $K\bar{K}$ pairs (left plot), time is measured in units of $10^{-10}$ seconds whereas for $B_d\bar{B_d}$ and $B_s\bar{B_s}$ pairs (middle and right plot), time is in units of $10^{-12}$ seconds. These reflect the approximate lifetimes of the particles. The bands in the left and middle plots represent the effects of decoherence, with decoherence parameter ($\lambda$) set to a $3\sigma$ upper bound. The right plot does not include such bands due to the lack of experimental evidence for $\lambda$ in $B_s$ mesons; in this case, $F_{\max}=E_F$.
	}
	\label{kbmeson}
\end{figure}



Typically, Bell's inequality is violated in these correlated meson systems for approximately half of the meson lifetime. We observe that the quantum correlations in these systems can be \textit{nontrivially different} from those in stable systems, as illustrated by the relationship between Bell's inequality violations and teleportation fidelity. A \textit{notable observation} is that the teleportation fidelity does not surpass the classical threshold value of 2/3 despite the violation of Bell's inequality. As in the case of neutrinos, the temporal quantum correlations can be studied by the LG inequality.
For the initial state $\ket{\bar{B}^o}$, the three measurement LG parameter takes the form~\cite{JK_SB_2018}:
\begin{equation}
	K_3 = 1 - 4 P_{\bar{B}^o \rightarrow B^o}(\Delta t) + 2P_{\bar{B}^o \rightarrow B^o}(2 \Delta t).
\end{equation}

\begin{figure}
\centering
	\includegraphics[width = 35 mm]{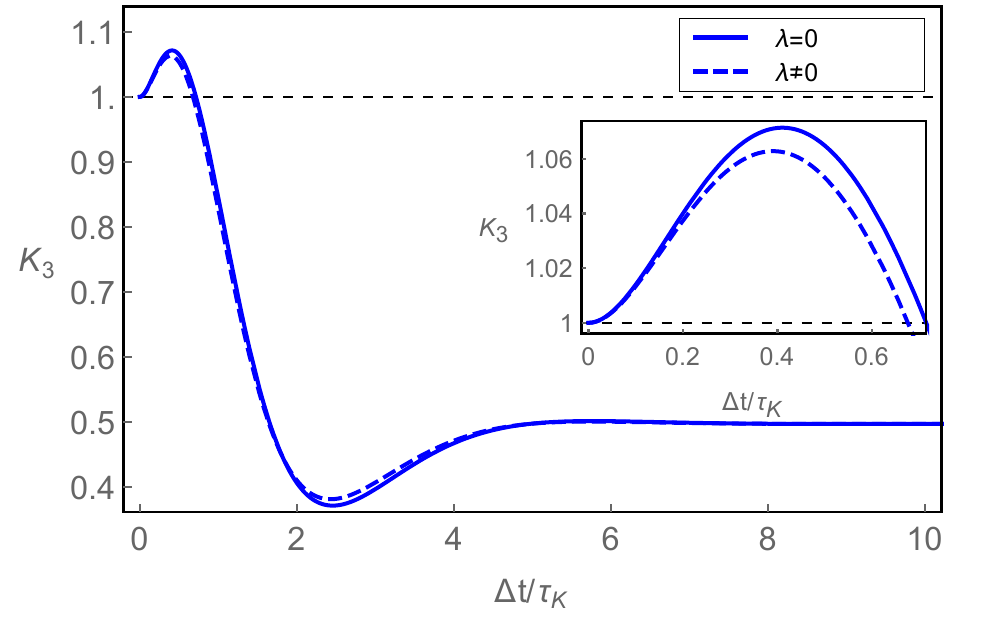}
	\includegraphics[width = 35 mm]{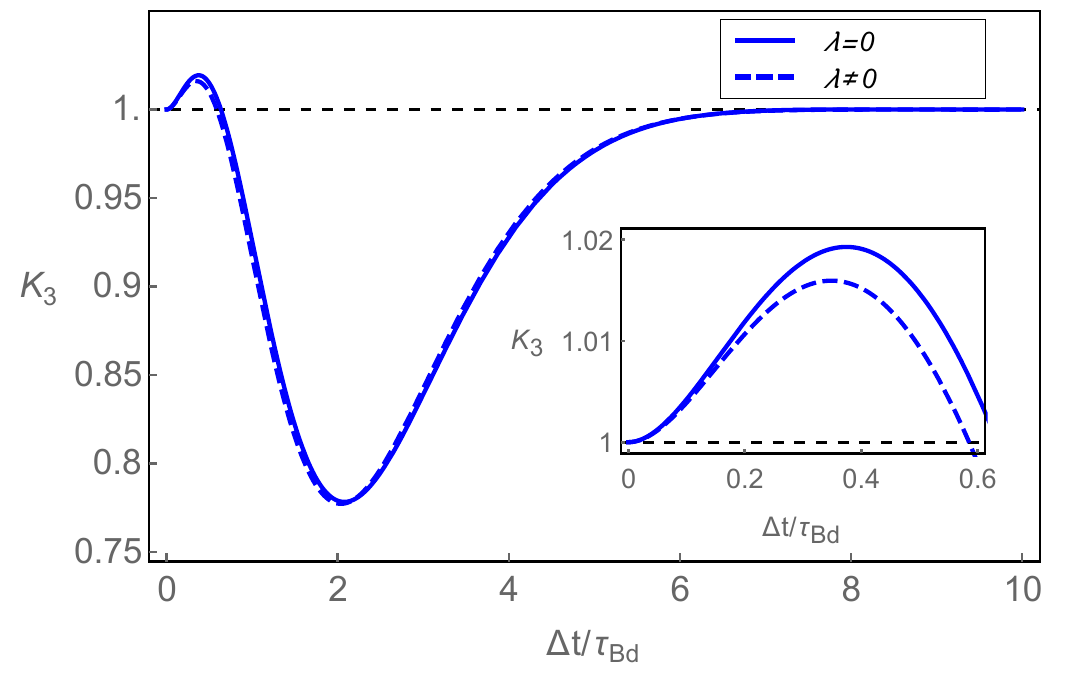}
	\includegraphics[width = 35 mm]{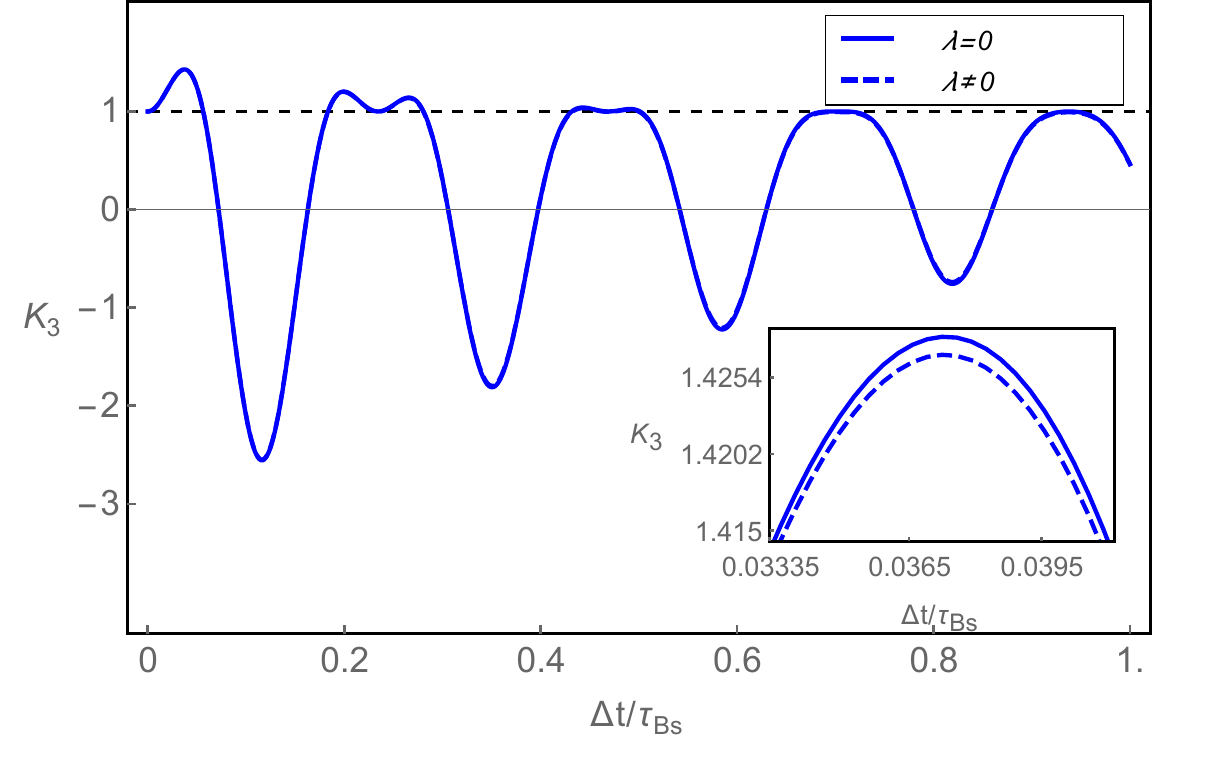}
	\caption{Depicting the LG parameter $K_3$ as a function of the time interval $\Delta t$ between the successive measurements. The left, middle, and right plots illustrate the $K$, $B_d$ and $B_s$ meson systems, respectively. The dashed line indicates the presence of decoherence, whereas the solid line represents the absence of decoherence. }	
 \label{fig_LG}
\end{figure}

As illustrated in Fig.~\ref{fig_LG}, the LG inequality is violated for some time intervals.
A knowledge of the transition probability at $\Delta t$ and $2\Delta t$ is needed to compute the function $K_3$.
\section{Conclusions}

We illustrated the role played by ideas in Open Quantum Systems and Quantum Information to various facets of neutrino and neutral meson oscillations.
The ubiquity of these tools is highlighted by the range of their applications, from quantum information to particle physics. The evolution of neutrinos was seen to be highly non-local in nature. The quantum correlations were closely tied to the the neutrino mixing energy. The correlations were shown to be simple functions of the product of neutrino survival and oscillation probabilities. The quantum correlations in the neutral mesons were seen to be non-trivially different from their stable counterparts.

\begin{acknowledgement}
I thank the organizers of ISRA 2023. My special thanks are due to Shubhrangshu Ghosh and Banibrata Mukhopadhyay. 
\end{acknowledgement}

\end{document}